\documentclass[twocolumn,showpacs,aps,amsmath,amssymb,nofootinbib]{revtex4-1}

\usepackage{color}   
\usepackage{graphicx}
\usepackage{dcolumn}
\usepackage{bm}

\begin{document}

\def\bb#1{\hbox{\boldmath${#1}$}}

\title{ Analytical Solutions of Landau (1+1)-Dimensional  Hydrodynamics }

\author{Cheuk-Yin Wong$^a$, Abhisek Sen$^b$, Jochen Gerhard$^c$, Giorgio Torrieri$^d$, Kenneth Read$^{ab}$}
\affiliation{$^a$Physics Division, Oak Ridge National Laboratory, Oak Ridge, TN  37831}
\affiliation{$^b$Department of Physics and Astronomy, University of Tennessee, Knoxville, TN  37996}

\affiliation{$^c$Frankfurt Institute for Advanced Studies (FIAS), Frankfurt am Main, Germany}

\affiliation{$^d$IFGW, Universidade Estadual de Campinas, Campinas, S$\tilde{a}$o Paulo, Brazil}

\date{\today}

\begin{abstract}

To help guide our intuition, summarize important features, and point
out essential elements, we review the analytical solutions of Landau
(1+1)-dimensional hydrodynamics and exhibit the full evolution of the
dynamics from the very beginning to subsequent times.  Special
emphasis is placed on the matching and the interplay between the
Khalatnikov solution and the Riemann simple wave solution, at the
earliest times and in the edge regions at later times.  These
analytical solutions collected and developed here serve well as a
useful guide and cross-check in the development of complicated
numerically-intensive relativistic hydrodynamical Monte Carlo
simulations presently needed.

\end{abstract}

\maketitle

\section{Introduction}

Landau hydrodynamics was put forth to study the dynamics of a
relativistic system possessing a simple equation of state in a
(1+1)-dimensional evolution \cite{Lan53,Bel55,Bel56,Bel65}.  The
accompanying Khalatnikov analytical solution is also well-known
\cite{Kha54} and has been discussed extensively in the literature.  It
forms the basis for many investigations in the rapidity distributions
and hydrodynamical behavior in high-energy heavy-ion collisions
\cite{Bel56a}-\cite{Mag05}.  Ref.\ \cite{Sri92} gives detailed
numerical results of the characteristics of the flow, $dN/dy$ as a
function of the freeze-out temperature, isotherms, and the difference
between the flow rapidity $y_{\rm flow}$ and the spatial rapidity
$y_s$, indicating the boost-non-invariance of Landau hydrodynamics.
Ref.\ \cite{Ham05} gives a detailed evolution of the entropy density
and temperature as a function of the longitudinal coordinate $z$ and
time $t$.  Numerical solutions have been presented earlier for (1+1)-
(2+1)- and (3+1)-dimensional hydrodynamics with the Landau initial
condition \cite{Ris96,Ris96a}.  Semi-analytic solution of the
(2+1)-dimensional hydrodynamics have been constructed by the method of
characteristics \cite{Bay83,Ham85}.  Numerous other elaborate numerical
calculations of relativistic hydrodynamics have been presented
\cite{Oll92,hydro,Sen14}.

While numerical hydrodynamical solutions serve well as tools for the
examination of the dynamics of many systems, the completely analytical
solutions remain useful to help guide our intuition, summarize
important features, and point out essential elements.  In this regard,
one finds three technical gaps for a completely analytical solution in
the existing literature.  First, conventional applications of Landau
hydrodynamics have been concentrated within the time domain under the
application of the Khalatnikov solution.  The Khalatnikov solution,
however, has its limitations.  It is not generally recognized that the
Khalatnikov solution is not applicable to discuss the hydrodynamics at
the earliest stages below a certain time coordinate.  We need to
specify an explicit analytical solution for the earliest history.
Secondly, even though the Khalatnikov solution is given in an
analytical form, the extraction of the solution is not as trivial as
it may appear to be.  An explicit procedure for the inversion of the
Khalatnikov solution from the space-time coordinates to the (energy
density)-velocity coordinates is needed.  Thirdly, even after the
Khalatnikov solution is inverted, only a part of the solution can be
utilized in the full hydrodynamical description.  As described in
\cite{Bel55,Bel56,Bel65,Kha54}, the Khalatnikov analytical solution
should be connected, in the vacuum side, to the Riemann simple wave
solution\footnote{For an exposition of the Riemann simple wave
  solution, see pages 366 and 503 of Landau and Lifshitz
  \cite{Lan58a}.}.  A complete hydrodynamical solution will need to
include the description of the matching transition and the connected
Riemann simple wave solution.  The present review has been motivated
to rectify the above gaps that hinders the application of the
analytical solutions of Landau hydrodynamics.

It should be pointed out that the earliest history of Landau
hydrodynamics is governed, not by the Khalatnikov solution, but by the
Riemann simple wave solution.  To obtain the full evolution dynamics,
we shall consider the initial Riemann simple wave solution and the
subsequent transitional matching of the Riemann simple wave solution
with the Khalatnikov solution.  In the discussions on the interaction
of jets with produced matter, which occur in the earliest stage of the
collision process, and on elliptic flows, which occur at the
subsequent early stage of hydrodynamical evolution, the early
hydrodynamics of the produced matter plays an important role and is of
considerable interest.  Furthermore, as hydrodynamics gains in
importance in high-energy heavy-ion collisions and
numerically-intensive hydrodynamics is being carried out with
supercomputers for multidimensional relativistic hydrodynamics on an
event-by-event basis \cite{Sen14}, simple analytical solutions will
provide great help in checking bench-mark results, guiding intuitions,
and comparing essential features, to ensure the success of the program
for our understanding of the hydrodynamical evolution process.

\section{The  Khalatnikov solution}

For the Landau initial condition of a reflectively symmetric slab of a
relativistic hot, dense matter initially at rest, the Khalatnikov
solution is an analytical solution of the hydrodynamical equation that
describes the space-time evolution of the system.  The solution is
obtained by introducing a hydrodynamical potential $\chi$ that is a
function of the energy density $\epsilon$ and the velocity $v$.  The
variables $\epsilon$ and $v$ can be alternatively represented by the
energy density logarithm $\zeta$ and the flow rapidity $y$,
\begin{eqnarray}
\zeta &=&
\frac{1}{4} \ln (\epsilon/\epsilon_0)= \ln (T/T_0),  \label{11}\\
\epsilon/\epsilon_0 &=& (T/T_0)^{4}=e^{4\zeta},
\label{22}\\\
s/s_0 = (\epsilon/\epsilon_0)^{3/4} &=& (T/T_0)^{3}=e^{3\zeta},\label{33}\\
v&=&\tanh y.\label{44}
\end{eqnarray}
Here $T$, and $s$ are the temperature and entropy density
respectively, and the subscripts $``0$'' denote initial values.  The
Khalatnikov solution consists of writing the space-time coordinates
$(z,t)$ as functions of $(\zeta,y)$ given (in
Eq.\ (4.12$'$) of \cite{Bel55}, Eq.\ (24) of \cite{Bel56}, and
Eq.\ (4.12a) of \cite{Bel65})  as
\begin{eqnarray}
t (\zeta,y)&=&e^{- \zeta } \left ( \frac{\partial \chi}{\partial  \zeta} \cosh y - \frac{\partial \chi}{\partial y }  \sinh y \right ),
\label{1}\\
x (\zeta,y)&=&e^{- \zeta } \left ( \frac{\partial \chi}{\partial  \zeta } \sinh y - \frac{\partial \chi}{\partial y }  \cosh y \right ).
\label{2}\end{eqnarray}
Belenkij and Landau considered a slab of width $2l$ initially at rest
and chose the origin of the longitudinal $z$-coordinate to be at
$x$=$-l$.  The longitudinal coordinate $z$ is therefore related to the
quantity $x$ in Eq.\ (\ref{2}) by
\begin{eqnarray}
z&=&x + l.
\end{eqnarray}
As we are considering a system possessing a reflection symmetry with
respect to $z$=0, we need to examine only the region of $z \ge$ 0.

The Khalatnikov solution is uniquely specified by the requirement to
satisfy two boundary conditions: (i) zero velocity ($v$=0 and $y$=0)
at the center of the symmetric slab at $z$=0 (and $x$=$-l$), and (ii)
the matching to the Riemann simple wave solution when $\zeta$=$-c_s y$
at the edge boundary of the slab.  In terms of the hydrodynamic
potential $\chi (\zeta,y)$, the Khalatnikov solution is given (in
Eq. (4.30) of \cite{Bel55,Bel65} and Eq.\ (26) of \cite{Bel56}) by
\begin{eqnarray}
\chi (\zeta,y)&=&-l \sqrt{3} e^\zeta \int_{y/\sqrt{3}}^{- \zeta } e^{2\zeta'} I_0\left [\sqrt{\zeta'^2 - \frac{1}{3} y^2}\right ] d\zeta'.
 \label{7}\end{eqnarray}
The above solution (\ref{7}) and the energy density relations in
Eqs.\ (\ref{11})-(\ref{33}) have been obtained for the equation of
state
\begin{eqnarray}
p=\frac{\epsilon }{3},
\label{8}
\end{eqnarray}
with the speed of sound 
\begin{eqnarray}
c_s=\sqrt{\partial p /\partial \epsilon}=1/\sqrt{3}.
\end{eqnarray}
We shall use the above speed of sound for our hydrodynamical
calculations.  The generalization of the analytical solutions of
Landau hydrodynamics to a general equation of state with a different
speed of sound $c_s$ can be found in Ref.\ \cite{Beu08} and is
summarized in Appendix A.

It is necessary to take note of the typographical errors in the
original articles of Belenkij and Landau \cite{Bel55,Bel56,Bel65} and
the change of notations.  The original Russian article in \cite{Bel55}
was presented in a simplified English version in \cite{Bel56} and in a
full English translation in \cite{Bel65}.  In conformity with the
standard notation to label the rapidity variable by $y$, we have
changed the notation of the rapidity variable $\alpha$ in
\cite{Bel55,Bel56,Bel65} to $y$ in Eq.\ (\ref{44}), and the energy
density logarithm variable $y$ in \cite{Bel55,Bel56,Bel65} to $\zeta$
in Eq.\ (\ref{11})-(\ref{33}).  To be consistent with Eqs.\ (\ref{1})
and (\ref{2}), the dimensionless energy density logarithm variable $y$
in the original articles of \cite{Bel55,Bel56,Bel65} should be defined
as $y$=$\ln(T/T_0)$ and not as $y$=$\ln T$.  The sign on the
right-hand side of the Khalatnikov solution, Eq.\ (4.30) in
\cite{Bel55,Bel65} and Eq.\ (26) in \cite{Bel56}, should be corrected
to be negative.  The factor preceding the integral in the Khalatnikov
solution should be $l\sqrt{3}e^y$ (as in \cite{Bel55} and
\cite{Bel56}), and not erroneously as $l\sqrt{3 e^y}$ as in
\cite{Bel65}.  The Khalatnikov solution Eq.\ (\ref{7}) in the present
article is the correct expression after all the typographical errors
have been corrected and the notations have been changed.

From an inspection of Eqs.\ (5), (6), and (8), it is clear that the
physical results of $t$ and $x$ are unchanged, if the right-hand sides
of Eqs.\ $\{$(5),(6),(8)$\}$ are multiplied by arbitrary constant
factors of $\{$$A$, $A$, $1/A\}$, respectively.   After these multiplications, the product $(A \chi)$ has
the same dimension as $x$ and $t$, namely, the dimension of 
length. 
Because of the
invariance of $t$ and $x$ with respect to different  choices of $A$, the
Khalatnikov solution can be written in many equivalent, and equally
valid, forms, with $A$=1 in 
 \cite{Bel55,Bel56,Bel65,Kha54,Bel56a,Mil59,Mil59a,Ros59,Gor85},
or $A$=$1/T_0$ in 
 \cite{Cha74,Sri92,Pai95,Beu08,Pes11,Bia11}.\
There is freedom in the choice of $A$ to 
partition the length dimension of $(A\chi)$ between $A$ and $\chi$, or
equivalently, to define $\chi$ in terms of  $t$ and $ x$ by writing  the Legendre transform equation (4.10) of
Belenkij and Landau
 \cite{Bel55,Bel65}  in a more general form with an explicit $T_0$ as
\begin{eqnarray}
d(A\chi) = d(\phi + \frac{T}{T_0}u^0 t - \frac{T}{T_0}u^1 x). 
\end{eqnarray} The original
Khalatnikov solution of Eqs.\ $\{$(5),(6),(8)$\}$ in
\cite{Bel55,Bel56,Bel65,Kha54,Bel56a,Mil59,Mil59a,Ros59,Gor85} corresponds to the choice of  $A$=1, requiring 
$\chi$ to carry the length dimension, whose scale turns out
to be $l$ in Eq.\ (8) as determined by the boundary condition of
$x$=$-l$ at $y$=0 for all $t$ [2-5].  Another choice selects a
dimensionless $\chi$, requiring  the factor $A$ to carry the length dimension, which
can be chosen to be the natural length scale of $x$ and $t$ with 
$A$=$l$, or the natural length scale associated with  $T$ with $A$=$1/T_0$.  The Khalatnikov solution
as expressed in  \cite{Cha74,Sri92,Pai95,Beu08,Pes11,Bia11} corresponds to
the choice of $A$=$1/T_0$, leading to equivalent, and equally valid, expressions obtained by
multiplying the right-hand sides of Eqs.\ $\{$(5), (6), (8)$\}$ by
$\{1/T_0$,$1/T_0$,$T_0\}$, respectively.  

\section{The Riemann Simple Wave Solution}

In the Khalatnikov solution in the last section, there are two
hydrodynamical degrees of freedom which have been chosen to be the
energy density $\epsilon$ and the velocity $v$, or alternatively,
$ (\zeta,y)$.  There is however another Riemann wave simple wave
solution of the one-dimensional relativistic hydrodynamical equations
in which the energy density represented by $\zeta$ and the velocity
represented by $y$ can be expressed as a function of each
other in which the space-time coordinates $x$ and $t$ do not
explicitly appear.  In the presence of a disturbance, the simple wave
propagation can be visualized as the superposition of (i) the
propagation of a sound wave with the speed of sound $c_s$ and (ii) the
propagation of the fluid element itself with a flow velocity $v=\tanh
y$.  They occur at the edge boundary regions where the energy density
decreases monotonically until the energy density vanishes, when the
matter is in contact with the vacuum.  As the two edge boundaries of the
slab are always in contact with the vacuum, the Riemann simple wave
solutions are always present on the slab boundaries.

Because of this mutual dependencies between $\zeta$ and $y$, there is
then only a single independent hydrodynamical degree of freedom in the
Riemann simple wave solution.  The hydrodynamics is described by
$\zeta(y(x,t))$ or vice-versa $y(\zeta(x,t))$ in the form of a running
wave whose profile can change with time.  In non-relativistic
hydrodynamics, the relation between the fluid density $\rho$ and the
velocity field $v$ in a simple wave are related by Eq. (94.4) of
Landau and Lifshitz \cite{Lan58a} :
\begin{eqnarray}
v  = \pm \int \frac{  dp}{c_s\rho}= \pm \int \frac{ c_s d\rho}{ \rho}.
\end{eqnarray}
This solution satisfies the equations of 1-D hydrodynamics.  In the
relativistic case, this becomes Eq. (2) on page 503 of Landau and
Lifshitz \cite{Lan58a}:
\begin{eqnarray}
y=  \tanh^{-1} v  
&=& \pm \int \frac{c_s d\epsilon  }{(\epsilon +p)}
= \pm \frac{1}{c_s } \ln \left \{ (\epsilon /\epsilon _0)^{c_s^2/(1+c_s^2)}\right \} ,
\nonumber
\end{eqnarray}
which leads to $y=\pm { \ln (T/T_0)}/{c_s }$ or
\begin{eqnarray}
 y&=& \pm \frac{ \zeta}{c _s}. 
\label{12}
\end{eqnarray}
The sign on the right-hand side of above equation is so chosen that it
gives the correct sign for $y$ (and $v$).  As the energy density
$\epsilon$ in general is less than $\epsilon_0$,
$\zeta$=$\frac{1}{4}\ln (\epsilon/\epsilon_0)$ is generally negative.
So, for the region of $z$$\ge$0 we are interested, we have $v > 0$ and
we should take the negative sign of ({\ref{12}).  Thus,
  Ref.\ \cite{Bel56} gives the condition for the simple wave as
\begin{eqnarray}
y = - \frac{\zeta}{c_s}.
\label{13}
\end{eqnarray}
In terms of the potential $\chi$ in Eqs. (\ref{1}) and (\ref{2}), we have
\begin{eqnarray}
\frac{x}{t}
&=&\frac{ \tanh y  - \frac{\partial \chi}{\partial y } /  \frac{\partial \chi}{\partial \zeta }}
{ 1-  \tanh y ~~(\frac{\partial \chi}{\partial y } /\frac{\partial \chi}{\partial \zeta }) }.
\end{eqnarray}
For simple waves with a center at the origin, the total derivative of
the potential function $\chi (\zeta,y)$ is zero \cite{Lan58a},
\begin{eqnarray}
\frac{d\chi (\zeta,y)}{d y}=\frac{\partial \chi (\zeta,y)}{\partial y}+\frac{\partial \chi (\zeta,y)}{\partial \zeta}\frac{d \zeta }{dy}=0.
\end{eqnarray}
So, we have 
\begin{eqnarray}
\frac{d \zeta }{dy} = -~{\frac{\partial \chi (\zeta,y)}{\partial y}}\biggl /{\frac{\partial \chi (\zeta,y)}{\partial \zeta}}.
\label{16a}
\end{eqnarray}
From Eq.\ (\ref{13}) and (\ref{16a}), we obtain
\begin{eqnarray}
\frac{\partial \chi}{dy } \biggl /  \frac{\partial \chi}{d \zeta }=- 
\frac{d \zeta }{dy} =
c_s,
\end{eqnarray}
and the Riemann simple wave solution  is 
\begin{eqnarray}
\frac{x}{t}=\frac {\tanh (-\zeta/c_s)-c_s}{1-\tanh (-\zeta/c_s) ~c_s}.
\label{18}
\end{eqnarray}
Eqs. ({\ref{13}) and (\ref{18}) constitute the Riemann simple wave solution
  for the edge boundary region of the slab.

\section{Early Hydrodynamical Evolution  at $t\le l/c_s$}

We consider the Landau initial condition of a full stopping resulting
in an initial slab of width $2l$ initially at rest, with an initial
energy density $\epsilon_0$, as shown in Fig.\ 1.  The slab is in
contact with the vacuum and the energy density of the slab decreases
monotonically, starting from the matter region to the vacuum region.
The hydrodynamical motion of the slab at the early moments is governed
by the Riemann simple wave solution specified by Eqs.\ (\ref{13}) and
(\ref{18}).

For a fixed value of $t$$\le$$l/c_s$, we increase the value of the
rapidity $y$ stepwise, starting from $y$=0.  Knowing the value of $y$,
we can calculate the energy density logarithm $\zeta$ from
Eq.\ (\ref{13}).  After obtaining $\zeta$, we can calculate $x$ from
Eq.\ (\ref{18}).  The calculation is repeated for the next value of
$y$.  As $y$ increases, $\zeta$ becomes more negative and the energy
density $\epsilon/\epsilon_0$=$e^{4\zeta}$ decreases until the density
becomes vanishingly small, and the velocity $v$ approaches 1.

The hydrodynamical solution at the early stage exhibits the following
features as shown in Fig 1.

\begin{figure}[h]
  \includegraphics[scale=0.45]{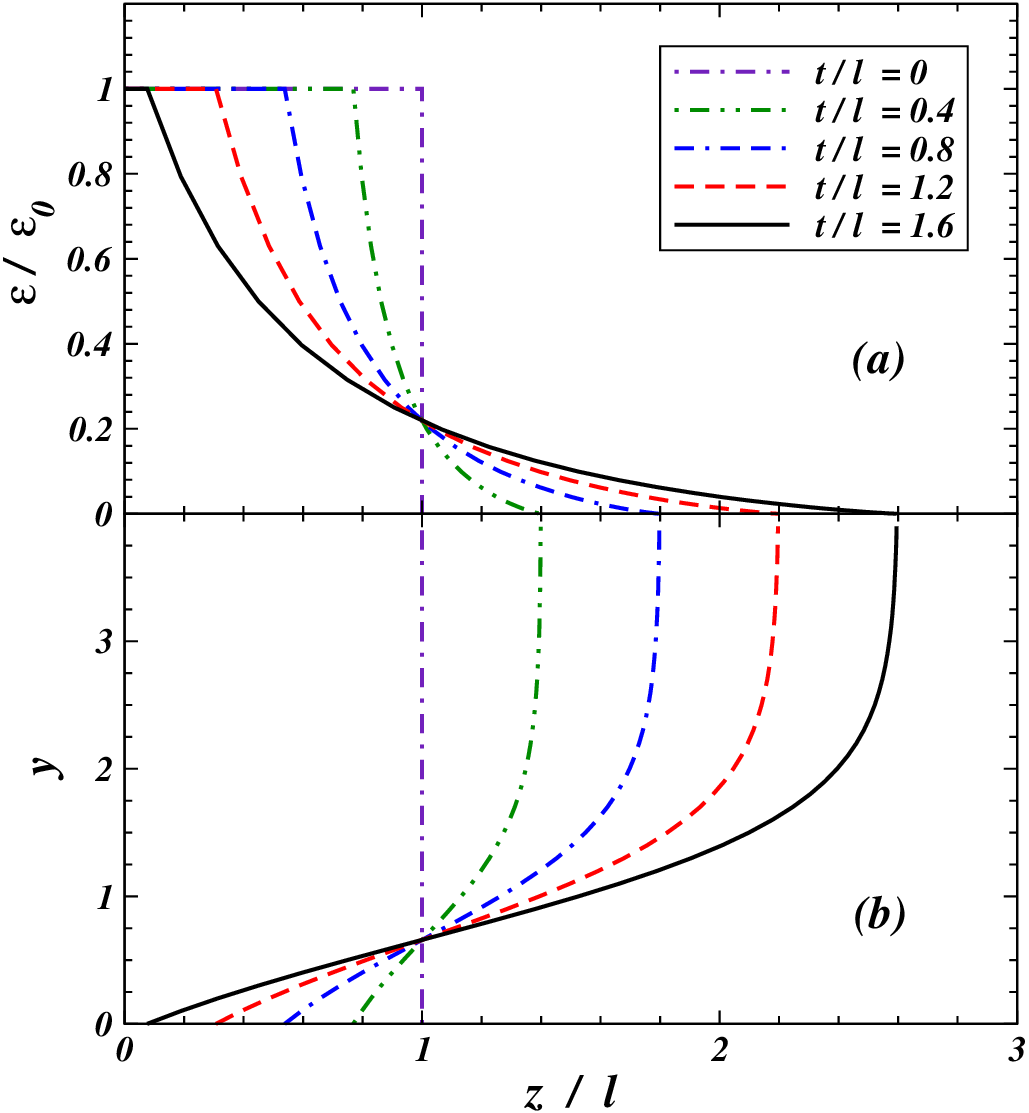}
  \caption{(Color online ) The ratio $\epsilon/\epsilon_0$ and flow
    rapidity $y$  as a function of $z/l$ for different values
    of $t/l $$\le$$ l/c_s$ obtained with the Riemann simple wave
    solution.}
\end{figure}

\begin{enumerate}

\item

For zero rapidity $y$=0 ($v$=0) with the fluid at rest, we have
$y$=0 ($\epsilon$=$\epsilon_0$) at the spatial coordinate $x$=$-c_s t$
(or $z$=$l$-$c_s t$).

The rarefaction wave starts at $z$=$l$ and propagates inward to $z$=0
with the speed of sound $c_s$.  The rarefaction wave reaches the
spatial origin $z$=0 at time $t=l/c_s$=$\sqrt{3}l$.

\item

As $y$  increases, $\zeta$ becomes more and more negative, and the
energy density $\epsilon$ decreases.  The variation of $y$  traces out
the whole curve of $\epsilon/\epsilon_0$ as a function of $z/l$ for a
fixed $t$.

\item

From Eq.\ (\ref{18}), we note that $x$=0 ($z$=$l$) occurs at $\zeta=- c_s
(\tanh ^{-1} c_s)$, for different times $t$.  Thus the curves of
$\epsilon/\epsilon_0$ for different $t$ meet at the same point of
$\epsilon/\epsilon_0 = \exp\{ -4 c_s
(\tanh ^{-1} c_s)\} \sim 0.22$ in Fig. 1.

\item
The fluid expands outward and the velocity of the fluid element
increases as the fluid coordinate increases.  The farthermost reach of
the fluid element occurs at $y $$\to$$ \infty$, ($v$$ \to$1 and $x$$
\sim$$t$), which corresponds to $z $$\sim $$(l+t$).  The velocities of
the fluid elements in contact with the vacuum approach to, and are
limited by, the speed of light.
 
\end{enumerate}

\section{ Hydrodynamic evolution after $ t \ge l/c_s$ }

After the time $t$$\ge$$l/c_s$, the rarefaction wave that starts from
the edge of the slab at $z$=$l$ reaches the center of the slab at
$z$=0 (Fig. 1).  Subsequent expansion of the fluid in the central
region will proceed through the Khalatnikov solution of
Eqs.\ (\ref{1}), (\ref{2}), and (\ref{7}).  To determine $ (\zeta,y)$
as a function of $(z,t)$, it is useful to express the derivatives of
$\chi (\zeta,y)$ explicitly in terms of $\zeta$ and $y$  so that
Eqs.\ (\ref{1}) and (\ref{2}) for the coordinates $(x,t)$ are explicit
functions of $ (\zeta,y)$.  The quantities $ (\zeta,y)$ can then be
inverted to become a function of $(x,t)$.

Using Eq.\ (8), we can take the derivative with respect to $\zeta$ and we get
\begin{eqnarray}
\frac{\partial \chi}{\partial \zeta} (\zeta,y) &=& \chi + 
l \sqrt{3} e^{- \zeta } I_0\left[\sqrt{\zeta^2 - \frac{1}{3} y^2}\right].
\label{19}\end{eqnarray}
We  take the derivative of $\chi$ with respect to $y$  and we get two terms,
\begin{eqnarray}
 \frac{\partial \chi}{\partial y}&=&-l \sqrt{3} e^\zeta  \int_{y/\sqrt{3}}^{- \zeta } e^{2\zeta'}\frac{\partial}{\partial y} I_0\left [\sqrt{\zeta'^2 - \frac{1}{3} y^2}\right ] d\zeta' +{\cal I}, \nonumber
\end{eqnarray}
where $\cal I$ is the derivative with respect only to the lower limit
$y/\sqrt{3}$.  We also have $I_0'(x)=I_1(x)$ \cite{Abr65}, and
thus
\begin{eqnarray}
 \frac{\partial \chi}{\partial y}&=&l(y/\sqrt{3}) e^\zeta  \int_{y/\sqrt{3}}^{- \zeta } e^{2\zeta'}\frac{  I_1\left [\sqrt{\zeta'^2 - \frac{1}{3} y^2}\right ]}{
{\sqrt{\zeta'^2 - \frac{1}{3} y^2}} }d\zeta'+{\cal I} .\nonumber
\end{eqnarray}
We can evaluate ${\cal I} $ to yield 
\begin{eqnarray}
{\cal I} 
&=&l  e^\zeta  e^{2 y/\sqrt{3}}.
\end{eqnarray}
Adding these two terms, we have
\begin{eqnarray}
 \frac{\partial \chi}{\partial y} (\zeta,y)&=&l (y/\sqrt{3})  e^\zeta  \int_{y/\sqrt{3}}^{- \zeta } e^{2\zeta'}
\frac{  I_1\left [\sqrt{\zeta '^2 - \frac{1}{3} y^2}\right ]}{
{\sqrt{\zeta'^2 - \frac{1}{3} y^2}} }d\zeta '
\nonumber\\
& &+l  e^\zeta  e^{2 y/\sqrt{3}}.
\label{21}\end{eqnarray}

With the knowledge of $\partial \chi/\partial \zeta$ and $\partial
\chi/\partial y$ given by Eqs.\ (\ref{19}) and (\ref{21}), the
right-hand sides of Eqs.\ (\ref{1}) and (\ref{2}) give $(x,t)$ as
explicit functions of $(\zeta, y)$.  The integral in Eq.\ (\ref{21})
can be evaluated numerically as the limits of the integration and the
integrands are known functions of $\zeta$ and $y$.

The hydrodynamical description is simplest if we succeed in expressing
$ (\zeta,y)$ as a function of $(z,t)$.  For this purpose, it is
necessary to invert Eqs.\ (\ref{1}) and (\ref{2}) from
$(z,t)$(function of $\zeta,y$) to $ (\zeta,y)$(function of $z,t$).  We
consider a fixed value of $t$, and we increase stepwise the value of
$y$, starting from zero.  For each pair values of $(t,y)$,
Eq.\ (\ref{1}) presents itself as an equation for the unknown quantity
$\zeta$ (or equivalently, $\epsilon/\epsilon_0$).  We can solve this
Eq.\ (\ref{1}) with only one unknown $\zeta$ by the Newton's method
using a good guessed value of $\zeta$, starting at $y=0$.  From
Eqs.\ (\ref{19}) and (\ref{1}), a good guess on the value of $\zeta$
for a given $t$ and $y$=0 is
\begin{eqnarray}
\zeta^{(0)}&=&-\frac{1}{2} \ln\left ( \frac{t}{\sqrt{3}l}\right ). 
\end{eqnarray}
Subsequent guesses can then be obtained using Newton's method after
numerically evaluating the change in the residue as a function of a
small change in $\zeta$.  Newton's method has a rapid convergence.
After the solution for $\zeta$ is obtained, Eqs.\ ({\ref{19}) and
  (\ref{21}) are then used with Eq.\ (\ref{2}) to calculate the value
  of $x$.  The newly determined $\zeta$ can be used as the guess for
  the next $y$ to get the new solution of $\zeta$.

\section {Khalatnikov solution and Matching to the Simple Wave solution for $t \ge l/c_s$}

\begin{figure}[h]
  \includegraphics[scale=0.45]{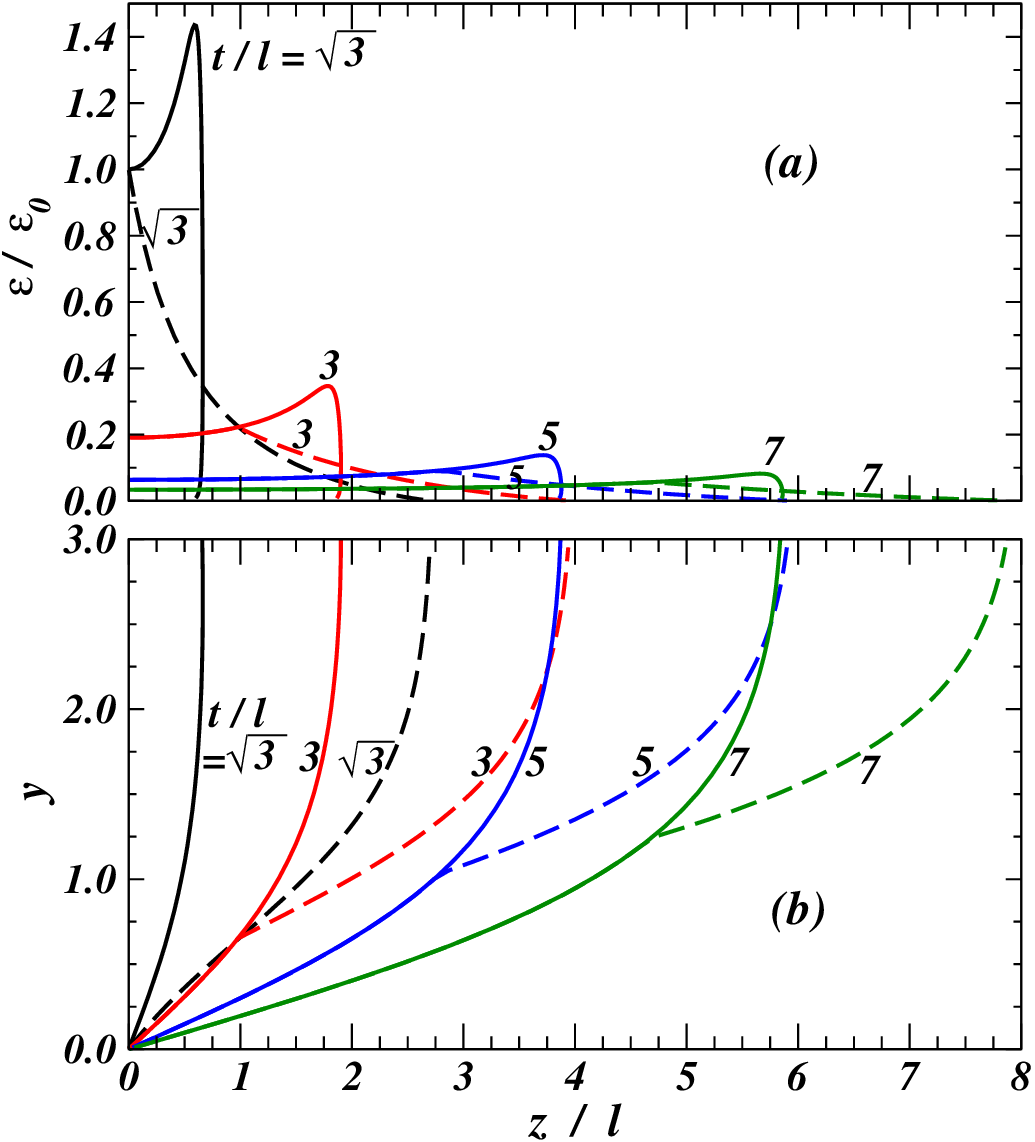}
  \caption{(Color online ) The quantities $(\epsilon/\epsilon_0,y)$ as
    a function of $z/l $ for different values of $t/l$.  The solid
    curves give the Khalatnikov solution which must be matched on to
    Riemann simple wave solutions at the edge boundaries shown as
    dashed curves.  A complete hydrodynamical solution consists of the
    Khalatnikov solution for small $z/l$ (solid curve) joining on to
    the matched Riemann solution for large $z/l$ (dashed curve).  }
\end{figure}

The Khalatnikov solution is not applicable before the time coordinate
$t< l/c_s$.  At $t$=$l/c_s$=$\sqrt{3}l$, the rarefaction wave has just
reached the center of the slab at $z=0$ and the fluid motion described
by the Khalatnikov solution has just started to become applicable.  We
show in Fig. 2 the Khalatnikov solution for $t/l$=1/$c_s$, 3, 5, and 7 as
a function of $z/l$ as solid curves.  At $t$=$l/c_s$, the Khalatnikov
solution has an energy density exceeding the initial density and
increasing as a function of $z$.  It decreases precipitously at $z/l$$\sim$0.7 .  At subsequent time coordinates, the energy density rises
as a function of $z$ and decreases precipitously near 
$t$$\sim $$(z-l)$.  The corresponding rapidity increases monotonically and rapidly
as a function of increasing $z/l$

However, not all portions of the Khalatnikov solution shown as the
solid curves can be used to describe the evolution of the system
because the dynamics at the edge region is described by the
propagation of a disturbance arising from the presence of the edge
boundary.  The accompanying hydrodynamical motion in the edge region
is a Riemann simple wave propagating from the edge toward the center.
The hydrodynamical solution at the edge of the slab is governed by the
Riemann simple wave solution.  The Khalatnikov solution that is
applicable in the interior of the slab needs to be matched on and
switched to the simple wave solution when the energy density logarithm
$y$ matches the rapidity $y$ by Eq.\ (\ref{13}), $y = -\zeta/c_s$.
For $t\ge l/c_s$, the complete hydrodynamical solution for the fluid
with the Landau initial condition consists of the Khalatnikov solution
in the interior region of small $|z|$, and the matched Riemann simple
wave solution at the edge boundaries of the system.
 
We can carry out the matching in the following way.  We study the
Khalatnikov solution for a fixed value of $t$ ($\ge$$ l/c_s$) and
increase stepwise the value of $y$, starting from $y$ =0.  We
calculate $\zeta$, $x$, and $z$ as a function of $t$ and $y$, using
the method outlined in the last section.  After determining $\zeta$
for the pair of $(t,y)$ values, we test whether
$-\zeta/c_s$=$-\sqrt{3} \zeta$ remains greater than $y$ or not.  If
$-\sqrt{3} \zeta$ remains greater than $y$, we proceed to the next
incremented value of $y$ and look for the Khalatnikov solution for the
next set of $(t,y)$ pair.  On the other hand, when $-\sqrt{3} \zeta$
is equal to or just begin to be greater than $y$, the hydrodynamical
solution will be switched from the Khalatnikov solution to the Riemann
simple wave solution for subsequent $y$ values.

For the Riemann simple wave solution in the boundary region for a
fixed value of $t$, we increase stepwise the value of $y$.  The energy
density logarithm variable $\zeta$ is then given by $\zeta=-c_s y$.
Knowing and the values of $t$, $y$ and $\zeta$, the spatial coordinate
$x$ is given by Eq.\ (\ref{18}).  This stepwise increase of $y$ allows
us to trace the energy density as a function of the longitudinal
coordinates.

\begin{figure}[h]
  \includegraphics[scale=0.45]{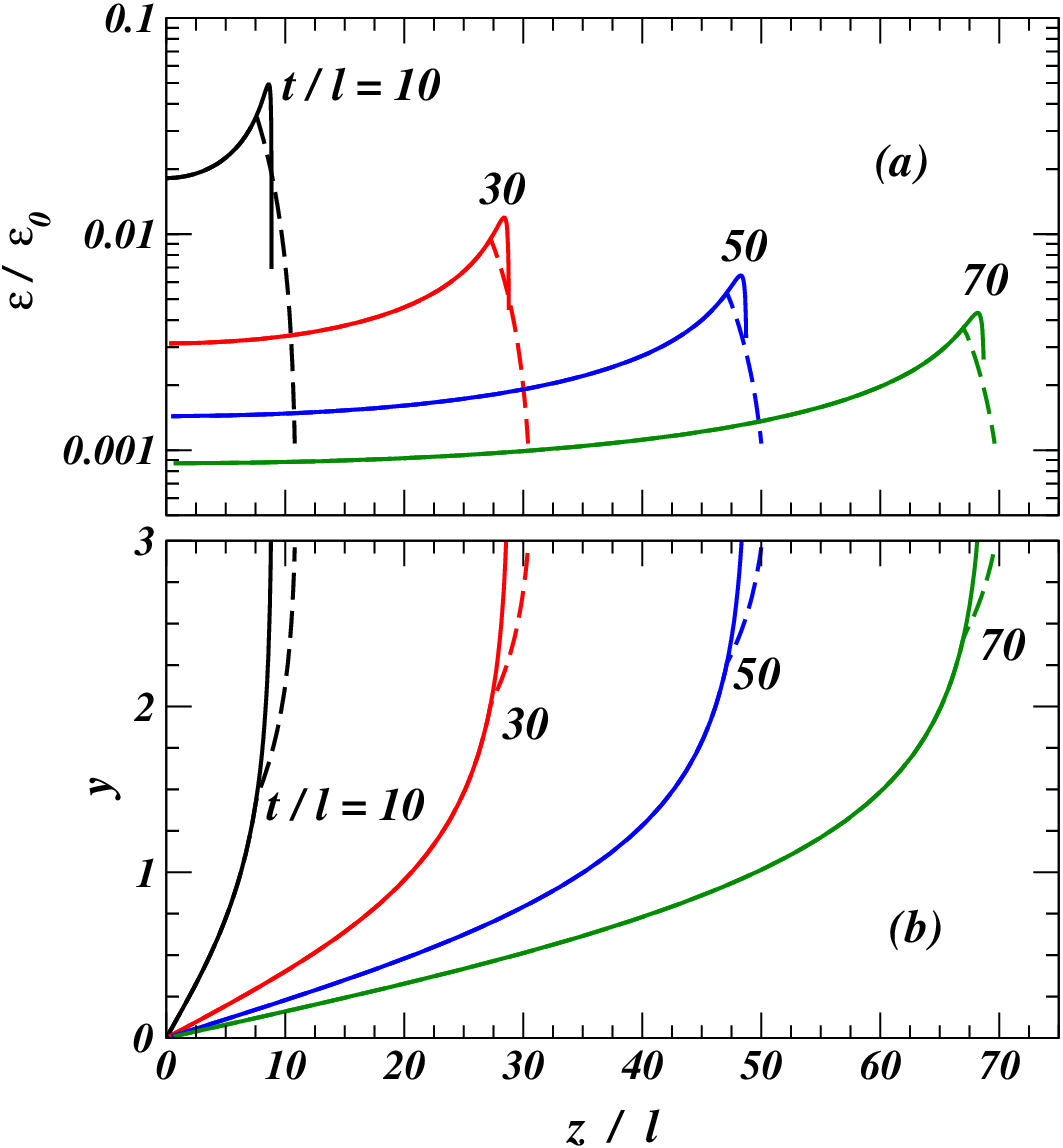} 
  \caption{(Color online ) The quantities $(\epsilon/\epsilon_0,y)$ as
    a function of $z/l $ for different values of $t/l$.  The solid
    curves give the Khalatnikov solutions which must be matched on to
    Riemann simple wave solutions shown as dashed curves. A complete
    hydrodynamical solution consists of the Khalatnikov solution  for small
    $z/l$ (solid curve) joining on to the matched Riemann solution for
    large $z/l$ (dashed curve). }
\end{figure}

At $t$=$l/c_s$=$\sqrt{3}l$, the matching of the Khalatnikov solution
with the Riemann simple wave solution occurs at $z=0$.  Thus the solid
curve of the Khalatnikov solution is not applicable at $t=\sqrt{3} l$.
In its place as the solution of Landau hydrodynamics is the Riemann
simple wave solution starting from $z=0$ shown as the dashed curve in
Fig. 2.  Therefore, at $t$=$l/c_s$=$\sqrt{3}l$, even though the
Khalatnikov solution begins to emerge, it does not contribute to the
hydrodynamical solution with the Landau initial condition.

At higher values of $t$, the fluid expands outward and the
longitudinal region under the Khalatnikov solution begins to expand.
At $t$=3$l$, the Khalatnikov solution extends to $z$$\sim$$l$ where
the matching with the Riemann simple wave solution occurs.  At $t=5l$,
the Khalatnikov solution extends farther out to $z$$\sim$$3 l$ where
the matching occurs.  The extension of the longitudinal region under
the Khalatnikov solution increases approximately linearly with time
$t$.  On the other hand, the extension of the Riemann simple wave
solution spans a longitudinal length of order $3l$ and is
approximately independent of $t$.  Thus, the Khalatnikov solution
covers a longitudinal region less than the Riemann waves for $t
\lesssim 5 l$, but a longitudinal region greater than the Riemann
waves for $t \gtrsim 5 l$.  The full hydrodynamical solution consists
of the Khalatnikov solution in the region of small $z$ (solid curves)
and the matched Riemann simple waves solution in the region of large
$z$ (dashed curves) in Fig. 2.  They are the hydrodynamical solutions
satisfying the boundary conditions.

We show in Fig. 3 the Khalatnikov solutions as solid curves for
$t/l$=10, 30, 50, and 70.  The Riemann simple wave solutions which match with
the Khalatnikov solutions are given as dashed curves.  At $t$=10$l$,
the Khalatnikov solution extends to $7.5 l$ and the simple wave solution 
extends over a length of about $3l$.  At later times when $t \gg  l$, the
matching occurs at a spatial coordinate just a few units less than $t$
with a simple wave that is approximately $3l$ in length.  As the
simple wave region extends approximately to only a few units of $l$
and $t \gg l$, the simple wave region is much smaller than the
Khalatnikov solution region for large values of $t$.

\section{Hydrodynamical solution in  ($\tau$, $y_s$) }

To study
the question of boost invariance, it is useful to
introduce  $\tau$ and $y_s$  which are   related to 
$(t,z)$ by
\begin{subequations}
\label{23}
\begin{eqnarray}
\tau&&=\sqrt{t^2-z^2} =\sqrt{t^2-(x+l)^2},\label{23a} \\
y_s&&=\frac{1}{2} \ln \frac{t+z}{t-z},\\
z&&=x+l .
\end{eqnarray}
\end{subequations}
The inverse relations are
\begin{subequations}
\label{24}
\begin{eqnarray}
t=\tau \cosh y_s,  \\
z=x+l = \tau \sinh y_s.
\end{eqnarray}
\end{subequations}
Strictly speaking, only for  solutions that are boost-invariant with respect to
the origin at $(t,z)$=0 can the quantity $\tau$ be properly called the proper
time and $y_s$ the associated spatial rapidity.  As we do not possess
a boost-invariant initial condition, the coordinates $(\tau,y_s)$ can
only be approximately and analogously identified with the proper time
and the spatial rapidity, respectively. Such an approximate
identification allows their use as tools to judge the degree of boost
invariance of a hydrodynamical evolution.  Specifically, at a constant
value of $\tau$, a boost-invariant hydrodynamical evolution will be
indicated by an energy density $\epsilon$ that is independent of $y_s$
and a flow rapidity $y$ equal to the spatial rapidity $y_s$.
Conversely, at a constant value of $\tau$, the deviation of $\epsilon$
from a constant as a function of $y_s$ or the inequality of $y$ and
$y_s$ will be an indication of boost-non-invariance.  The degree to
which $\tau$ can be approximately identified as the proper time will
depend on how close to boost invariance the solution will turn out to
be.

With this choice of the $(\tau,y_s)$ coordinates, only regions with
$t$$>$$|z|$ possess real $\tau$ and $y_s$ to fall within our realm of 
description.  The limits of real $\tau$ and $y_s$ are the straight
lines $t$=$\pm z$ for which $\tau$=0.  Therefore, at all times $t$,
there are boundary edge regions with a finite width $\Delta z$=$l$ in
the simple wave regions, for which $|z|$$>$$t$, and $\tau$ and $y_s$
are not real.  Such small edge boundary regions fall outside our realm of  
description.  

We need to express the Khalatnikov solution and the Riemann solution
in terms of  $\tau$ and  $y_s$.  We
represent $(\tau,y_s)$ in terms of $(t,x)$ by Eq.\ (23) which are in
turn represented as functions of $\zeta$ and $y$ by Eqs.\ (\ref{1})
and (\ref{2}), with the hydrodynamical potential $\chi$ determined by
Eq.\ (\ref{7}).  With the knowledge of $\partial \chi/\partial \zeta$
and $\partial \chi/\partial y$ given by Eqs.\ (\ref{19}) and
(\ref{21}), the $(\tau, y_s)$ variables are explicit functions of
$(\zeta,y)$.

We can express the Riemann simple wave solution as a function of the
$(\tau, y_s)$ coordinates by substituting (\ref{24}) into (\ref{18}).
We introduce the effective velocity $a$ as
\begin{eqnarray}
x=\tau \sinh y_s -l&=& a \tau \cosh y_s =at,
\end{eqnarray}
where
\begin{eqnarray}
 a=\frac{v-c_s}{1-vc_s}.
\end{eqnarray}
The Riemann simple wave solution in terms of $(\tau,y_s)$ becomes
\begin{eqnarray}
e^{y_s}&=& \frac{1 + \sqrt{1 + (\tau/l)^2 (1-a^2)}}{(\tau/l)(1-a )} ,
\label{27}
\end{eqnarray}
which describes the hydrodynamical motion of disturbances in the
boundary regions of the slab.  The Khalatnikov solution needs to match
to the Riemann simple wave solution when the energy density logarithm $\zeta$
and the rapidity $y$ are related by the speed of sound as $\zeta=- c_s
y$ given by Eq.\ (\ref{13}).

\begin{figure}[h]
  \includegraphics[scale=0.45]{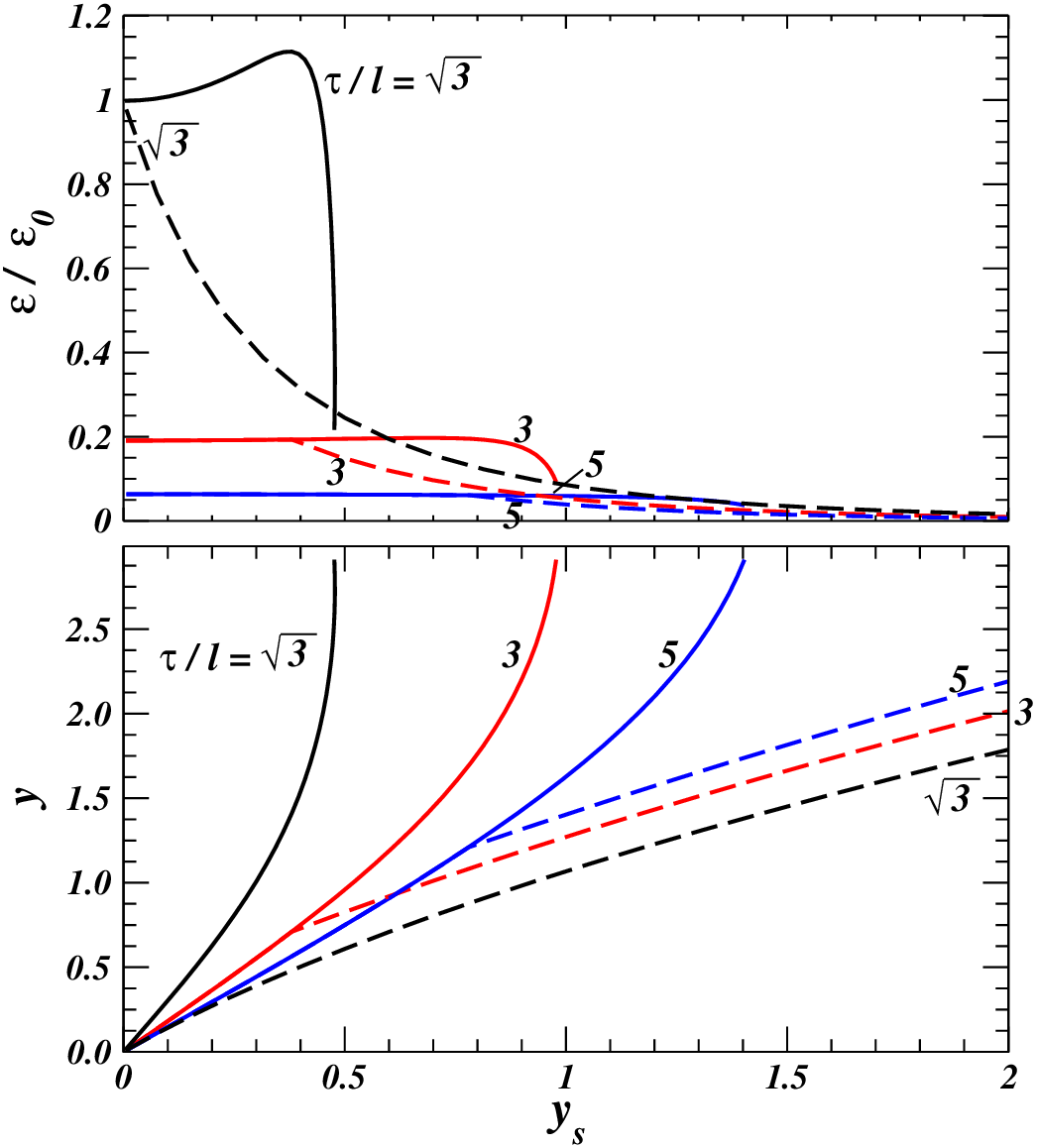}
  \caption{(Color online ) The quantities $(\epsilon/\epsilon_0,y)$ as
    a function of the spatial rapidity $y_s $ for different values of
    $\tau/l$.  The solid curves give the Khalatnikov solutions which
    must be matched on to Riemann simple wave solutions shown as
    dashed curves.  A complete hydrodynamical solution consists of the
    Khalatnikov solution for small $y_s$ (solid curve) joining on to the
    matched Riemann solution for large $y_s$ (dashed curve).  }
\end{figure}

 We consider a fixed value of $\tau$$\ge$$l/c_s$ and stepwise increase
 the value of $y$, starting from $y$ =0.  We can obtain the
 hydrodynamical description of $ (\zeta,y)$ as a function of $(\tau,
 y_s)$ by inverting Eq.\ (\ref{23}) and its associated equations.  For
 each pair of $(\tau,y)$ values, equation (\ref{23a}) together with
 the associated supplementary equations (\ref{1}) and (\ref{2})
 presents itself as an equation for the unknown quantity $\zeta$.  We
 can solve this equation with only one unknown $\zeta$ by Newton's method
 using a satisfactory guessed value of $\zeta$, starting at $y$=0.
 From Eqs.\ (\ref{19}) and (\ref{1}), a good guess (trial value) for
 the value of $\zeta$ at $y$ =0  for a given $\tau$ is
\begin{eqnarray}
\zeta^{(0)}&=&-\frac{1}{2} \ln\left ( \frac{\tau}{\sqrt{3}l}\right ). 
\end{eqnarray}
After the solution of $\zeta$ is obtained, Eqs.\ ({\ref{19}) and
  (\ref{21}) are then used with Eq.\ (\ref{2}) to calculate the value
  of $x$, $z$ and $y_s$.  The newly determined $\zeta$ can be used as
  the trial value for the next $y$ to get the new solution of $\zeta$.

\begin{figure}[h]
  \includegraphics[scale=0.45]{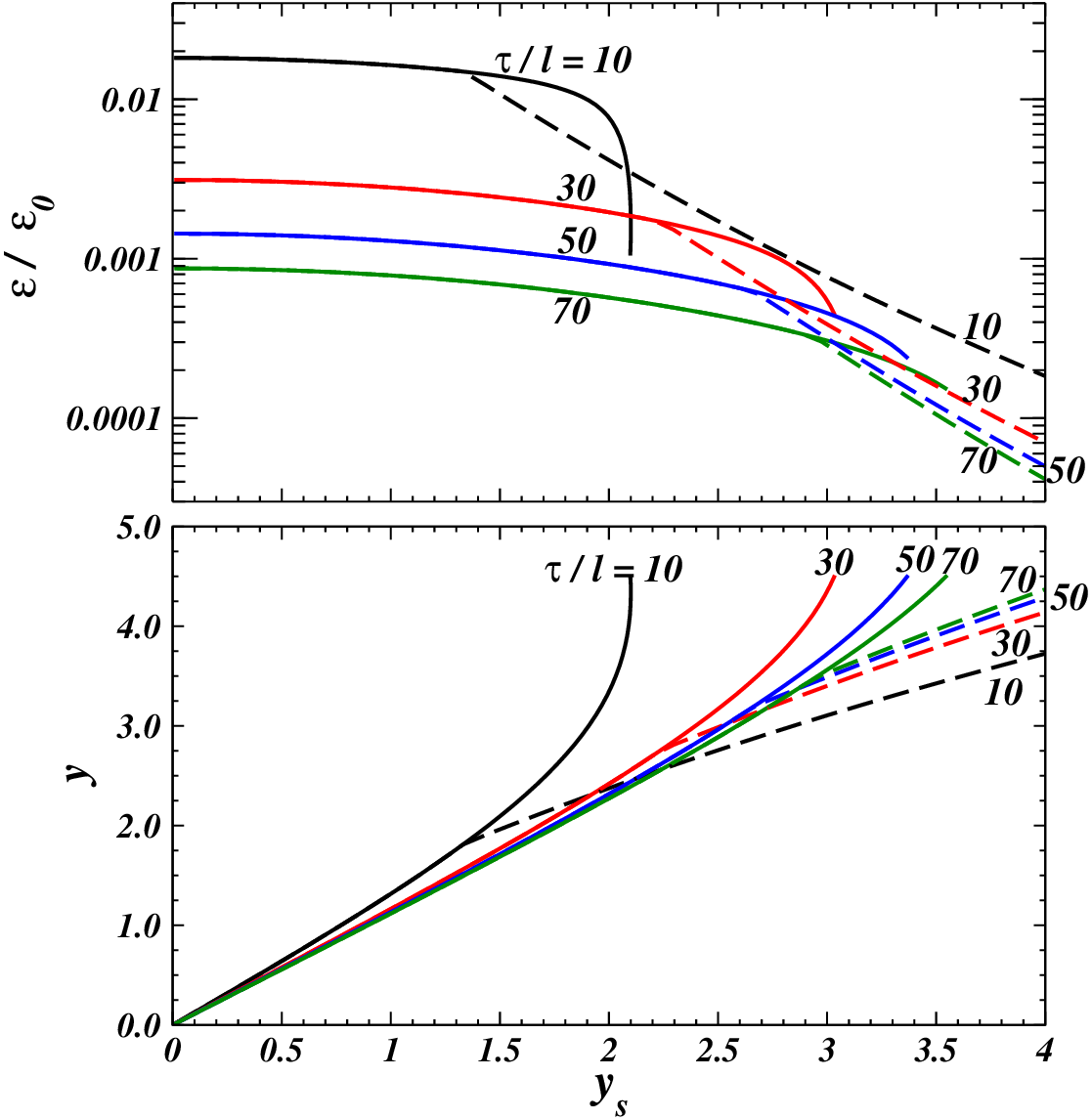}
  \caption{(Color online ) The quantities $(\epsilon/\epsilon_0,y)$ as
    a function of the spatial rapidity $y_s $ for different values of
    $\tau/l$.  The solid curves give the Khalatnikov solutions which
    must be matched on and switched to the Riemann simple wave
    solutions shown as dashed curves.  }
\end{figure}

Using the method we have just outlined for $\tau$$\ge$$l/c_s$, we can
determine the Khalatnikov solution as a function of the spatial
rapidity $y_s$ for a fixed value of $\tau$ shown as solid curves in
Fig. 4. 
In the time domain of Fig.\ 4, $((\epsilon/\epsilon_0)$ is
relatively flat as a function of $y_s$ but the flow rapidity $y$ is
consistently greater than $y_s$ except at very large values of $y_s$.
 However, not all parts of the Khalatnikov solution can be
used for our complete hydrodynamical solution.  It is necessary to
match the Khalatnikov solution to the Riemann simple wave when $\zeta$
is equal to $-c_s y$.  

  We carry out the simple wave
matching of the Khalatnikov solution by testing $-\zeta$ against $c_s
y$.  When $-\zeta $ is equal to or just begin to be greater than $c_s
y$, the solution will be switched to the Riemann simple wave solution
for subsequent $y $ values.  For this Riemann simple wave solution,
the energy density logarithm variable $\zeta$ is given by $\zeta =-c_s
y$ and the spatial rapidity $y_s$ is given by Eq.\ (\ref{27}).  The
complete hydrodynamical solution consists of the Khalatnikov solution in the
region of small $y_s$ (solid curves) and the matched Riemann simple
waves solution in the region of large $y_s$ (dashed curves) in Fig. 4.

We show in Fig. 5 the dynamics of the system for later times of
$\tau=10$ 30, 50, and 70 $l$.  In this time domain, the energy density
in the region of small $y_s$ decreases as a function of $y_s$.  For
example, for $\tau/l$=70, $(\epsilon/\epsilon_0)$ decreases by a
factor of three as $y_s$ increases from 0 to 3, indicating a lack of
boost invariance for this value of $\tau$.  The flow rapidity $y$ is
slightly greater than the spatial rapidity $y_s$.

\section{Other comparisons}

The solution of $ (\zeta,y)$ as a function of $(t,z)$ or $(\tau,y_s)$
allows one to extract other hydrodynamical quantity of interest.  In
addition to the energy density, one can calculate the spatial profiles
of the temperature or entropy density at different times $t$ or proper
times $\tau$.

\begin{figure}[h]
  \includegraphics[scale=0.65]{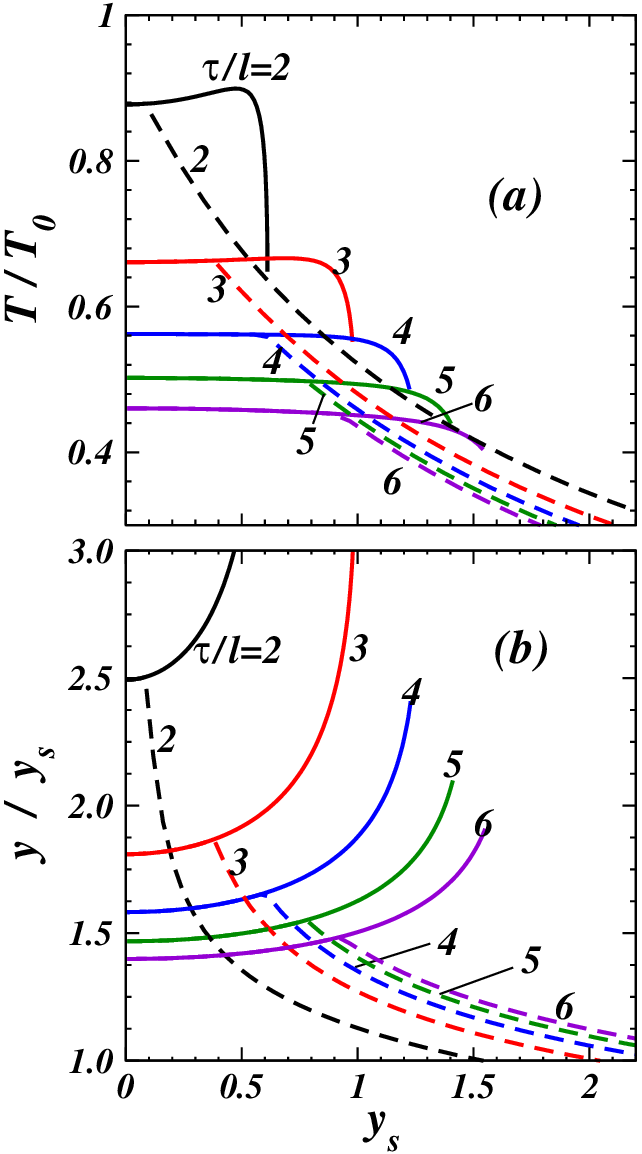}
\caption{(Color online ) The quantities $T/T_0$ and $y/y_s$ as a
  function of $y_s$ for different values of the $\tau/l$.
  The solid curves give the Khalatnikov solutions for small $y_s$
  which must be matched on and switched to the Riemann simple wave
  solutions for large $y_s$ shown as dashed curves.}
\end{figure}

We show the ratio $T/T_0$ in Fig.\ 6a, and the ratio $y/y_s$ in
Fig.\ 6b, as a function of $y_s$ for different values of the proper
time $\tau/l$.  We observe that for small values of $\tau/l$= 2-6, the
Khalatnikov solution starts to emerge from the central region, the
longitudinal length of the Khalatnikov solution included into the
hydrodynamical description gradually increases.  In this time domain,
the temperature or the energy density of the Khalatnikov solution is
relatively flat as a function of $y_s$, but the ratio $y/y_s$ is
consistently greater than unity, which indicates a high degree of
boost-non-invariance, especially during the early stage of the
hydrodynamical evolution.

\begin{figure}[h]
  \includegraphics[scale=0.65]{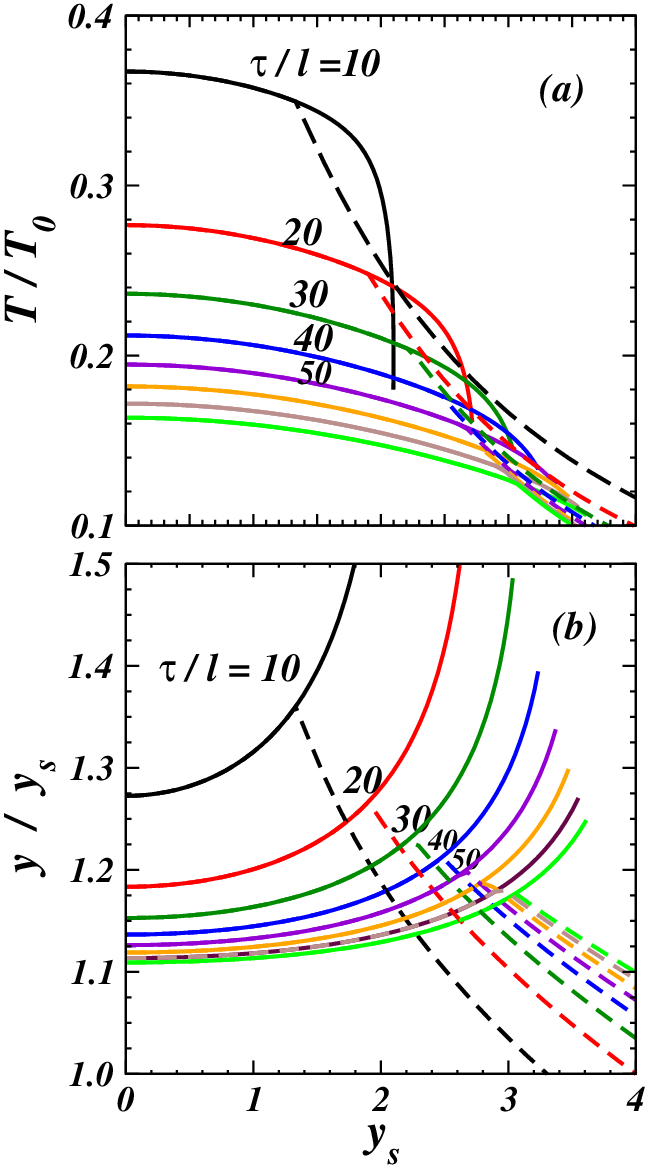}
\caption{(Color online ) The quantities $T/T_0$ and $y/y_s$ as a
  function of $\tau/l $ for different values of the spatial rapidity
  $y_s$.  The solid curves give the Khalatnikov solutions for small
  $y_s$ which must be matched on to Riemann simple wave solutions for
  large $y_s$ shown as dashed curves.}
\end{figure}

Fig.\ 7 gives $T/T_0$ and $y/y_s$ as a function of $y_s$ for different
 $\tau/l$=10, 20, 30, 40, and 50.  We observe that the
temperature decreases gradually as a function of the spatial rapidity
$y_s$, and the ratio of $y/y_s$ is consistently greater than unity,
even for $\tau$$\sim $ 80$l$.

\begin{figure}[h]
  \includegraphics[scale=0.40]{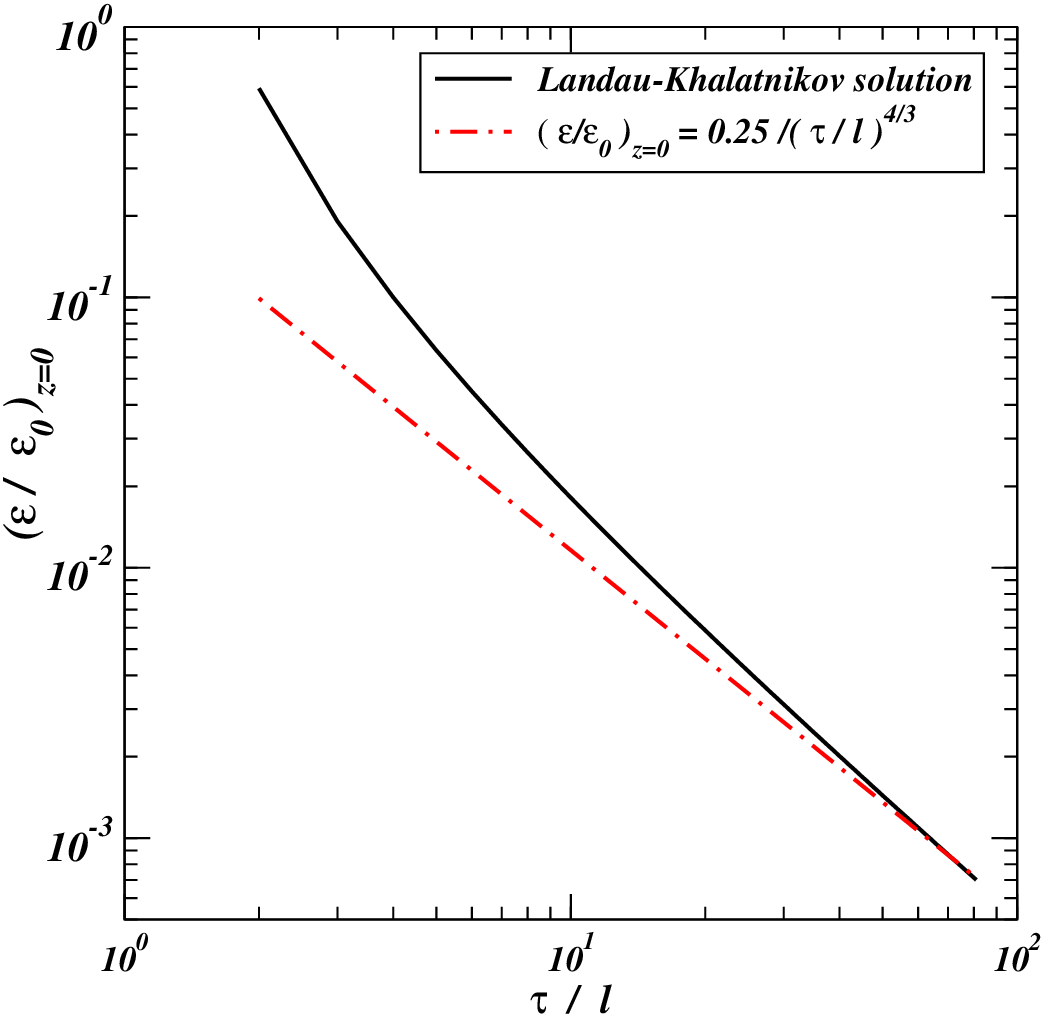}
\caption{(Color online ) The ratio $\epsilon/\epsilon_0$ at $z$=0 ($y_s$=0) as a
  function of $\tau /l $.  The solid curves represent the solutions
  from the Khalatnikov solution with the Landau initial condition, and
  the dashed curve is the behavior expected from Bjorken
  hydrodynamics.}
\end{figure}

In Fig.\ 8, we show the ratio $\epsilon/\epsilon_0$ at the center of
the slab at $z$=0 as a function of the $\tau /l$.  The
energy density decreases with $\tau/l$ but the decrease does not
follow the Bjorken limit of $\epsilon/\epsilon_0 \propto
1/\tau^{4/3}$.  Bjorken-like behavior of $\epsilon/\epsilon_0 \propto
1/\tau^{4/3}$ behavior occurs only at the very late stage of $\tau /l
\sim$ 80.

The relation between $\epsilon/\epsilon_0$ and $\tau/l$ at $z$=0 ($y_s$=0) in
Fig.\ 8 can be fitted very well by the empirical formula
\begin{equation}
\label{epsilonimp}
\frac{\epsilon}{\epsilon_0} \simeq b\left( \frac{\tau}{l}\right)^{-\frac{4}{3} + a \frac{l}{\tau} + c \left( \frac{l}{\tau} \right)^d}
\end{equation}
where, respectively, $a \simeq 2.60, b\simeq 0.213,c \simeq 2.25, d
\simeq 3.48$.  This above formula can be used to provide an effective
correction to the equivalent Bjorken energy formula
\begin{equation}
\frac{\epsilon}{\epsilon_0}= b' \left(\frac{\tau}{l }\right)^{-4/3} 
\end{equation}
that is usually used to estimate the initial energy density given the
initial time $\tau_0$ and the experimentally observed energy density
$dE/dy$, which, for a rapidity-independent system is
\begin{equation}
\label{epsilonexp}
\epsilon = \frac{1}{ S \tau_0}\frac{dE}{dy} ,
\end{equation}
where $S$ is the transverse area of the system.  Since the Landau
model does not require full stopping but just lack of transparency
(see the introduction in \cite{Sen14}), the initial energy density
compatible with the Landau model is not necessarily $\sqrt{s}
N_{part}$ (which, at ultra-relativistic energies is too high).  The
initial energy density assuming a Landau initial condition can instead
be estimated from Eqs. (\ref{epsilonimp}) and (\ref{epsilonexp})
scaled by $y_s/y$, given an estimate of the initial time $\tau_0$ of
the system.

Similarly, Fig.\ 7 is well-described at $z$=0 ($y_s$=0), also in the
asymptotic limit $\lim_{\tau \rightarrow \infty} y/y_s$=1, by this
parametrization
\begin{equation}
\frac{y}{y_s} \simeq 1 +a \left(  \frac{l}{\tau} \right)^b +c \left( \frac{l}{\tau} \right)^d,
\end{equation}
where $a=1.942,b=1.178,c=0.220,d=0.184$.  This parametrization can be
used to obtain a back-of-the-envelope estimate of the goodness of the
Bjorken approximation, assuming Landau initial conditions and a given
initial time $\tau_0$ that is approximately related to the initial
slab width by $\tau_0 \sim l$.

Transverse expansion will of course alter these approximation to
$\mathcal{O}(50\%)$, but for that a realistic numerical calculation
such as \cite{Sen14} is required.

\section{Conclusions and Discussions}

We undertake our present review to rectify three technical gaps that
hinders the application of the analytical solutions of Landau
hydrodynamics.  First, we show that the earliest history can be
described exclusively by the Riemann simple wave solution.  Secondly,
the inversion of the Khalatnikov solution can be carried out
successfully with well-outlined procedures.  Thirdly, we show how the
Khalatnikov solution and the Riemann simple wave solution can be matched
at different time domains.  In consequence, the analytical Khalatnikov
solution and the matched Riemann simple wave solution provide a
complete picture of the full evolution of the relativistic
hydrodynamics of a (1+1)-dimensional system.  Our examination with the
Landau initial condition reveals that the Riemann simple wave
solutions are always present at the two edge boundaries of the slab,
and the Khalatnikov solution properly appears only after the time
coordinate $t$$\ge$$l/c_s$.

The evolution can be depicted as following three stages of
development.  In the first stage of $t$$\le$$ l/c_s$, a Riemann
simple wave (rarefaction wave) moves towards the center and depletes
the density near the central region.  One edge of the simple wave
reaches the center of the slab at $t$=$l/c_s$.  The other edge expands
the matter into the vacuum.  In the edge region of matter expansion, the
velocity increases with the distance from the center, and the matter
always approaches the speed of light as it comes in contact with the
vacuum.  In this first stage, the Riemann simple wave solution
suffices to describe the hydrodynamical evolution.

At the second stage of $\sim$$5 l/c_s$$\gtrsim$$t$$\ge$$ l/c_s $, the
interior region begins to expand, and both the Riemann solution and
the Khalatnikov solution occupy comparable longitudinal regions and
must be used simultaneously in different longitudinal regions to
describe the hydrodynamical evolution.  Such a situation arises
because the Khalatnikov solution describes only the hydrodynamical
evolution of the system in the interior region whereas the dynamics at
the edge is described by the propagation of a disturbance arising from
the presence of the boundary edge.  The accompanying hydrodynamical
motion is a Riemann simple wave propagating from the edge boundary
toward the center.  The longitudinal length of the hydrodynamical
motion governed by the Khalatnikov solution and the Riemann simple
wave solution depends on the time in the Khalatnikov solution
expansion, $t-l/c_s$.  The greater is the time $t-l/c_s$ compared to
the Riemann simple wave characteristic time $\sim 2l/c_s$, the greater
is the spatial region governed by the Khalatnikov solution.

In the third stage when $t$$\gtrsim$$( \sim$$6l/c_s $), the
hydrodynamical motion is dominated by the Khalatnikov solution, with
the simple waves occupying only a relatively small longitudinal region
at the boundary edges.  The Khalatnikov solution suffices
approximately for the description of the hydrodynamics of the system,
if the edge boundary region can be neglected.  While the Khalatnikov
and the simple wave interplay at different stages of the
hydrodynamical evolution, Belenkij and Landau showed that entropy of
the system is concentrated in the central region while total energy
 (including both internal and kinetic energies and as seen in
the laboratory frame) is concentrated in the boundary region
\cite{Bel56}.

As hydrodynamics gains in importance in high-energy heavy-ion
collisions, the method of extracting the analytical solutions
presented here may be useful for those who would like to use the
procedure to examine the approximate behavior of a relativistic system
undergoing a one-dimensional expansion.  In fact, as shown earlier by
Rischke and Gyulassy \cite{Ris96,Ris96a}, the main features of the
hydrodynamics of (2+1)- and (3+1)-dimensional relativistic
hydrodynamics contains many features similar to the (1+1)-dimensional
system.  An explicit outline presented here on how the different
analytical solutions interplay in a completely analytical treatment
enhances our understanding of the hydrodynamical evolution process.

With regard to the question of the comparison of Landau hydrodynamics
and Hwa-Bjorken boost-invariance hydrodynamics \cite{Hwa74,Bjo83}, we
note that boost invariance implies that not only is the energy density
independent of $y_s$, the flow rapidity $y$ should also coincide with the
spatial rapidity $y_s$.  As shown previously in Landau (1+1)-dimensional
hydrodynamics in \cite{Sri92} and in numerically-intensive
event-by-event (3+1)-dimensional hydrodynamics with supercomputers
\cite{Sen14}, the Landau initial condition does not possess boost
invariance and during the Landau hydrodynamical evolution the flow
rapidity does not equal the spatial rapidity even at late times.  The
approach to boost-invariance appears to be a slow process, even though
the energy density or temperature appears to be relatively flat as a
function of $y_s$ \cite{Sen14}.

  This research used resources of the Oak Ridge Leadership Computing
 Facility at the Oak Ridge National Laboratory, which is supported by
 the Office of Science of the U.S.  Department of Energy under
 Contract No. DE-AC05-00OR22725.  GT also acknowledges support from
 DOE under Grant No. DE-FG02-93ER40764.

\vspace*{0.6cm}

\appendix

\section{ Generalization of the analytical solutions of Landau hydrodynamics for different  $c_s$}

For completeness, we summarize below 
the analytical solutions of Landau hydrodynamics 
and the dependencies on 
 the speed of sound $c_s$.  We consider an
equation of state
\begin{eqnarray}
p=c_s^2 \epsilon, 
\end{eqnarray}
where $c_s$ is assumed to be a constant.  The relations between
the energy density, entropy density, and the temperature in
Eqs.\ (\ref{11})-(\ref{33}) are modified to be
\begin{eqnarray}
\zeta &=& \ln (T/T_0)=
\frac{c_s^2}{1+c_s^2} \ln (\epsilon/\epsilon_0), 
\\
\epsilon/\epsilon_0 &=& (T/T_0)^{1+1/c_s^2}=e^{\zeta(1+c_s^2)/c_s^2},
\\
s/s_0& =& (\epsilon/\epsilon_0)^{1/(1+c_s^2)} = (T/T_0)^{1/c_s^2}=e^{\zeta/c_s^2}.
\end{eqnarray}
The space-time coordinates $(t,x)$ are related to $(\zeta,y)$
and the hydrodynamical potential $\chi$ as in Eqs.\ (\ref{1}) and
(\ref{2}).  When the speed of sound $c_s$ is taken into account, the Khalatnikov solution, Eq.\ (\ref{7}),  is modified to be  \cite{Beu08}
\begin{eqnarray}
\chi(\zeta,y)\!=\!
 - \frac{le^\zeta }{c_s} \!\! \int^{-\zeta}_{c_s y}  \!\!e^{ \frac{c_s^2+1}{2c_s^2} \zeta'}
\! I_0\!\left (\frac{1-c_s^2}{2c_s^2}\sqrt{\zeta'^2-{c_s^2 y^2}}\right )  \! d \zeta'.~~~~~~
\end{eqnarray}
The Riemann solution as a function of the speed of sound $c_s$ is already given by Eqs.\ (\ref{13}) and (\ref{18}).

\end{document}